\documentclass[prd,aps,a4paper,eqsecnum,nofootinbib,floatfix,twocolumn]{revtex4}  %
\newif\ifusesec
\usesectrue  
   
\usepackage{graphicx} 
\usepackage{amsmath,amsfonts,amssymb}
\usepackage{color}
\newcommand{\beq}{\begin{equation}}
\newcommand{\eeq}{\end{equation}}
\newcommand{\bea}{\begin{eqnarray}}
\newcommand{\eea}{\end{eqnarray}}

\begin{document}

\title{High-post-Newtonian-order dynamical effects induced by tail-of-tail interactions \\ in a two body system}

\author{Donato Bini$^{1}$, Thibault Damour$^2$, Andrea Geralico$^{1}$}
  \affiliation{
$^1$Istituto per le Applicazioni del Calcolo ``M. Picone,'' CNR, I-00185 Rome, Italy\\
$^2$Institut des Hautes Etudes Scientifiques, 91440 Bures-sur-Yvette, France
}

\date{\today}

\begin{abstract}
Starting from the recently derived conservative tail-of-tail action 
[D. Bini and T. Damour,  arXiv:2504.20204 [hep-th]] we compute several dynamical observables of binary systems (Delaunay Hamiltonian, scattering angle), at the  6.5 post-Newtonian accuracy  and up
to the 8th post-Minkowskian order. We find perfect agreement with
 previous self-force results, and (when inserting a recent high-post-Newtonian order derivation of
 radiated angular momentum [A. Geralico, arXiv:2507.03442 [gr-qc]]) with state-of-the-art  post-Minkowskian 
 scattering results [M.~Driesse et al., 
Nature \textbf{641}, no.8063, 603-607 (2025)].
\end{abstract}

\maketitle

\section{Introduction}

In recent years the study of the gravitational two-body problem has been enriched by 
combining, in a synergistic way, different theoretical approaches: post-Newtonian (PN) \cite{Blanchet:2013haa}, post-Minkowskian (PM) \cite{Damour:2016gwp,Damour:2017zjx,Bini:2017xzy,Vines:2017hyw,Bjerrum-Bohr:2019kec,Kalin:2020mvi,Damour:2019lcq,Mogull:2020sak},
self-force (SF) \cite{Barack:2018yvs}, effective-field-theory (EFT) \cite{Goldberger:2004jt,Foffa:2013qca,Porto:2016pyg,Foffa:2021pkg}, effective one-body (EOB) \cite{Buonanno:1998gg,Damour:2000we}, Tutti Frutti (TF) \cite{Bini:2019nra,Bini:2020hmy,Bini:2020nsb,Bini:2020wpo,Bini:2020rzn}.  
The comparison between these various approaches can be
done by computing gauge-invariant observables (such as the four-momentum changes of the individual bodies during a scattering encounter), or gauge-fixed quantities (such as the EOB Hamiltonian in a specific gauge).

Starting at the fourth PN order, and the fourth PM one, i.e., at $O(\frac{G^4}{c^8})$, the dynamics of
binary systems involves non-local-in-time interactions mediated by tail effects \cite{Blanchet:1987wq,Foffa:2011np,Galley:2015kus}.
The conservative part of dynamical tail effects is described by a time-symmetric non-local-in-time
action \cite{Damour:2014jta}  of the form
\bea
\label{eqtail0}
S_{\rm time-sym}^{\rm tail}&=& \frac{G{\mathcal M}}{c^3}  \int dt \times\nonumber \\
&&  {\rm Pf}_{2 \frac{ r_{12}(t)}{c}} \int  \frac{dt'}{|t-t'|} {\cal F}^{\rm split}(t,t')\,.
\eea
Here, ${\mathcal M}\equiv E/c^2$ is the total mass-energy of the system, and ${\cal F}^{\rm split}(t,t')$ is the time-split version of the 
instantaneous energy flux (see \cite{Bini:2020wpo} for details, notably about the definition
of the Partie finie operation, Pf, linked to the logarithmic divergence of the $t'$ integral at $t'=t$).

Starting at $O(\frac{G^5}{c^{11}})$, i.e., at the 5.5PN order, and the fifth PM one,  the dynamics of
binary systems also involves non-local-in-time interactions mediated by tail-of-tail effects. The
conservative part of tail-of-tail effects is described by a non-local-in-time action whose
leading-order contribution (due to the even-parity quadrupole radiation) has been derived in 
Refs. \cite{Damour:2015isa,Bini:2020wpo}, and whose complete expression (involving the
exchange of all, even- and odd-parity, multipoles) has been recently derived 
in two different forms. The manifestly time-symmetric form of the tail-of-tail action
involves the logarithmic kernel $\ln \frac{c|t-t'|}{2r_0}$ (where $r_0$ is an arbitrary scale which actually does not contribute to the action) and reads
\bea
\label{eqtailoftail0}
S_{\rm time-sym}^{\rm tail-of-tail}&=&\frac12 \left(\frac{G{\mathcal M}}{c^3}\right)^2G \sum_{l\ge 2}\frac{1}{c^{2l+1}} a_l \beta_l^{\rm even}\int dt \times \nonumber\\
&&\int_{-\infty}^\infty dt' I_L^{(l+2)}(t)I_L^{(l+2)}(t')\ln \frac{c|t-t'|}{2r_0}\nonumber\\
&+& \frac12 \left(\frac{G{\mathcal M}}{c^3}\right)^2G \sum_{l\ge 2} \frac{1}{c^{2l+3}} b_l \beta_l^{\rm odd}\int dt \times \nonumber\\
&&\int_{-\infty}^\infty dt' J_L^{(l+2)}(t)J_L^{(l+2)}(t')\ln \frac{c|t-t'|}{2r_0}\,,\nonumber\\
\eea
Here, $I_L$ and $J_L$ are the even- and odd-parity source multipole moments (as defined in the Multipolar Post-Minkowskian formalism), and $a_l$ and $b_l$ denote the coefficients entering the even and odd multipolar parts of the energy flux~\cite{Blanchet:2013haa}
\bea
a_l&=&\frac{(l+1)(l+2)}{l(l-1)l! (2l+1)!!}\,,\nonumber\\ 
b_l&=&\frac{4 l (l+2)}{(l-1)(l+1)! (2l+1)!!}\,.
\eea
In addition  $\beta_l^{\rm (even,odd)}$ are the beta function coefficients describing the scale dependence of the multipole moments. $\beta_l^{\rm (even)}$ was first computed in the Appendix of Ref. \cite{Blanchet:1987wq}; its meaning  as a coefficient  related to scale-dependence was then pointed out in   Ref. \cite{Blanchet:1997jj} and understood as a renormalization-group $\beta$ coefficient in Ref. \cite{Goldberger:2009qd}. The coefficient $\beta_l^{\rm (odd)}$ was computed in Refs. \cite{Almeida:2021jyt,Fucito:2024wlg,Ivanov:2025ozg}. The authors of Ref. \cite{Almeida:2021jyt} have recently 
(see the erratum, and the  arxiv version 3, of Ref. \cite{Almeida:2021jyt}) reconsidered and corrected their original result,
and so confirmed the result of \cite{Fucito:2024wlg,Ivanov:2025ozg}, namely the fact that the odd-parity $\beta_l^{\rm (odd)}$'s are equal to their even-parity counterparts $\beta_l^{\rm (even)}$.

By contrast with the linear-tail conservative nonlocal action \eqref{eqtail0} which is logarithmically UV-divergent 
(and which is completed, when considering the radiation-reacted dynamics, by a corresponding  UV-finite nonlocal, time-antisymmetric  radiation-reaction term; see
Sec. VI of \cite{Damour:2014jta}), the
 quadratic-tail conservative nonlocal action, Eq. \eqref{eqtailoftail0}, is UV-finite at $t'=t$. When considering
  the radiation-reacted dynamics it must be completed by a corresponding  (in-out) multipolar radiation-reaction term which is logarithmically divergent. [The latter, external-zone, logarithmic UV-divergence will be compensated by corresponding IR-divergences in the matched source multipole moments.]
  
 The main aim of the present work is to derive  several different observable quantities involving the tail-of-tail
 action, Eq. \eqref{eqtailoftail0}. These quantities will be derived at the absolute 6.5 PN accuracy, i.e., the
 fractional 1PN accuracy beyond the leading 5.5 PN order ($O(\frac{G^5}{c^{11}}))$. Concerning the PM
 accuracy, we will go up to the 7PM order ($O(G^7)$)  for elliptic motions, and up to the 
  8PM order ($O(G^8)$)  for hyperbolic motions. For elliptic motions, we shall derive the averaged
 (``Delaunay")  tail-of-tail Hamiltonian (together with its EOB transcription), while, for hyperbolic motions, we shall derive
  the scattering angle. These observables will  provide new results, and also checks both of the
  tail-of-tail action (in particular direct checks of  some of the beta coefficients, including the 
  odd-parity ones \cite{Fucito:2024wlg,Ivanov:2025ozg})
  and of recent high-accuracy computations \cite{Driesse:2024feo,Heissenberg:2025ocy,Geralico:2025rof}.

 \section{Time-split form of the tail-of-tail action}
 
 It will be convenient for our explicit tail-of-tail computations to use a \lq\lq time-split" reformulation 
 (involving  a scale-free principal-value integral) of the
 manifestly time-symmetric action \eqref{eqtailoftail0}, namely
\bea
\label{eqtailoftail}
S^{{\rm tail-of-tail}}_{\rm time-split}&=& \frac14  \left(\frac{G{\mathcal M}}{c^3}\right)^2G \int dt \int _{-\infty}^\infty \frac{d \tau}{\tau} {\mathcal G}^{\rm split}(t,\tau)\,,\qquad
\eea
with
\bea
\label{actionbetagen}
{\mathcal G}^{\rm split}(t,\tau)&=&\sum_{l\ge 2}\frac{1}{c^{2l+1}}\left( a_l \beta_l^{\rm (even)}{\mathcal G}_I^l(t,\tau)\right.\nonumber\\
&+&\left. \frac{b_l}{c^2} \beta_l^{\rm (odd)}{\mathcal G}_J^l(t,\tau)\right)\,,
\eea 
where
\bea
{\mathcal G}_I^l(t,\tau)&\equiv &I_L^{(l+1)}(t)[I_L^{(l+2)}(t+\tau)-I_L^{(l+2)}(t-\tau)]\,, \nonumber\\
{\mathcal G}_J^l(t,\tau)&\equiv & J_L^{(l+1)}(t)[J_L^{(l+2)}(t+\tau)-J_L^{(l+2)}(t-\tau)]\,.\qquad
\eea
In other words, introducing (both for electric and magnetic multipoles $X$) the shorthand notation
\beq
{\sf G}_X^l=\int dt \int_{-\infty}^\infty \frac{d\tau}{\tau} {\mathcal G}_X^l(t,\tau)\,.
\eeq
we have
\bea
\label{eqtailoftail2}
S^{{\rm tail-of-tail}}_{\rm time-split}&=& \frac{G}4  \left(\frac{G{\mathcal M}}{c^3}\right)^2 
\sum_{l\ge 2}\frac{1}{c^{2l+1}}
\left( a_l \beta_l^{\rm (even)}{\sf G}_I^l\right.\nonumber\\
&+&\left. \frac{b_l}{c^2} \beta_l^{\rm (odd)}{\sf G}_J^l\right)\,.\qquad
\eea
Working at the fractionally 1PN level, and using $\beta_l^{\rm (odd)}=\beta_l^{\rm (even)}=\beta_l$, we 
obtain
\bea
\label{eqtailoftail3}
S^{{\rm tail-of-tail}}_{\rm time-split}&=& \frac{G}4  \left(\frac{G{\mathcal M}}{c^3}\right)^2 \frac{1}{c^5}\left[ 
\left( a_2 \beta_2 {\sf G}_{I_2}
+\frac{b_2}{c^2} \beta_2 {\sf G}_{J_2}\right)\right.\nonumber\\ 
&+&\left.
\frac{1}{c^2}a_3 \beta_3 {\sf G}_{I_3}\right]+\ldots\,,
\eea
in which one should insert the explicit values
\beq
a_2=\frac15\,,\quad a_3=\frac{1}{189}\,,\quad b_2=\frac{16}{45}\,,
\eeq
and  
\beq
\beta_2=-\frac{214}{105}\,,\quad \beta_3=-\frac{26}{21}\,.
\eeq

\section{Tail-of-tail averaged Hamiltonian, and EOB Hamiltonian, along ellipticlike orbits at the 6.5 PN  accuracy}
\label{secelliptic}

A characteristic feature of the tail-of-tail conservative dynamics is to contribute at half-integer
PN orders, starting with the 5.5PN order ($O(\frac1{c^{11}})$). When considering tail-of-tail
dynamical effects at the 1PN fractional accuracy, the tail-of-tail action is the only contributor
of  5.5PN and 6.5PN dynamical effects.
Using first-order self-force (1SF) techniques (notably Mano-Suzuki-Takasugi (MST) black hole perturbation 
theory \cite{Mano:1996vt},
and the first law of binary black hole mechanics \cite{LeTiec:2011ab}),
Refs.  \cite{Bini:2014nfa,Bini:2016qtx}  computed the 1SF dynamical effects at the  5.5PN and 6.5PN
levels\footnote{In fact,  Refs.  \cite{Bini:2014nfa,Bini:2016qtx}  reached the 9.5 PN accuracy.}
by computing two (related) physical observables: the averaged 5.5PN + 6.5PN Hamiltonian,
and the corresponding EOB Hamiltonian, as parametrized by some half-integer contributions to
the effective metric, namely the $O(\nu)$ contributions to the EOB potentials 
$A(u,\nu)= - g_{00}^{\rm eff}$, ${\bar D}(u,\nu)=( - g_{00}^{\rm eff} g_{rr}^{\rm eff})^{-1}$
and $Q(u, p_r,\nu)$ (parametrizing contributions at least quartic in $p_r$ in the 
Damour-Jaranowski-Sch\"afer gauge we use).

In this Section, we compute the averaged tail-of-tail Hamiltonian following from the action 
\eqref{eqtailoftail0}, working only up to quadratic order in eccentricity, and 
we transcribe our results in EOB form by computing the corresponding contributions
to $A(u,\nu)= - g_{00}^{\rm eff}$, and ${\bar D}(u,\nu)=( - g_{00}^{\rm eff} g_{rr}^{\rm eff})^{-1}$.
[We would need to work at least to fourth order in eccentricity to derive the tail-of-tail
contributions to the   $Q$ potential.]
This computation will allow us to get non trivial checks of the tail-of-tail action by comparing
the results directly derived from the 6.5PN-accurate action \eqref{eqtailoftail3} to the 1SF results
of Refs.  \cite{Bini:2014nfa,Bini:2016qtx} obtained from black hole perturbation theory.

The starting point of our computation is the 6.5PN-accurate time-domain expression \eqref{eqtailoftail3}.
[We recall  the 1PN expression for the ADM mass of the system:
${\mathcal M}=M\left(1 - \frac{\nu}{2a_r}\eta^2\right)$, where $\eta \equiv \frac1c$.]
To compute its time average we describe the ellipticlike dynamics  by using the quasi-Keplerian parametrization of the motion (at the 1PN level of accuracy; $u$ denoting here the eccentric anomaly) 
\bea \label{quasiKeplerian}
 nt&=&\ell=u-e_t\sin u\,,\nonumber \\
 r&=&a_r(1-e_r\cos u)\,,\nonumber \\
\phi&=&2K\arctan \left[\left( \frac{1+e_\phi}{1-e_\phi} \right)^{1/2}\tan \frac{u}{2}  \right]\,. 
\eea
Here,  $n=2\pi/P$ with $P$  the period of the radial motion and $K=\Phi/2\pi$ denotes the periastron advance: both $n$ and $K$ are gauge-invariant quantities. 
On the other hand,
the various eccentricity parameters, $e_t \equiv e_t^h$, $e_r \equiv e_r^h$ and $e_\phi \equiv e_\phi^h$, and the semi-latus rectum, $a_r \equiv a_r^h$,  are coordinate-dependent; here, we use (suppressing the
superscript $h$ for brevity) their values in harmonic coordinates,  which can be invariantly defined by their (1PN-accurate) 
values in terms of the energy, $\bar E\equiv (H-M)/\mu$, and angular momentum, $j \equiv J/(G M \mu)$,
 of the system (see, e.g., Table III in Ref. \cite{Bini:2020wpo}). We also use  (dimensionless)
mass-rescaled  radial variables,  $r \equiv r^{\rm phys}/(G M)$, $a_r \equiv a_r^{\rm phys}/(G M)$.

The building blocks needed to evaluate the tail-of-tail action \eqref{eqtailoftail3} along ellipticlike orbits
are the explicit expressions of the bilinears in multipolar moments entering the tail-of-tail split action (PN-expanded, eccentricity expanded and expressed as functions of the eccentric anomaly $u$),
namely  $I_{ij}^{(3)}(t(u))I_{ij}^{(4)}(t'(u'))$, $J_{ij}^{(3)}(t(u))J_{ij}^{(4)}(t'(u'))$ and $I_{ijk}^{(4)}(t(u))I_{ijk}^{(5)}(t'(u'))$.
They are listed in Table \ref{tab:1} of Appendix  \ref{mult_mom}.

As already stated above, we limit ourselves to the 1PN fractional accuracy,
and we work at the second order in the (small) eccentricity expansion.
As in previous works, we use as basic variables the time eccentricity, $e_t$ and 
the radial semi-major axis $a_r$.

The tail-of-tail Hamiltonian is related to the corresponding action in the usual way
\beq
S^{{\rm tail-of-tail}}_{\rm time-split}=-\int dt H^{{\rm tail-of-tail}}(t)\,.
\eeq
Therefore
\bea
H^{{\rm tail-of-tail}}(t)&=&H^{{\rm tail-of-tail}}_{I_2}(t)+\eta^2 [H^{{\rm tail-of-tail}}_{J_2}(t)\nonumber\\
&+& H^{{\rm tail-of-tail}}_{I_3}(t)]\,,
\eea
where
\begin{widetext}
\bea
H^{{\rm tail-of-tail}}_{I_2}(t)&=& -\frac{G}{4}\left( \frac{G{\mathcal M}}{c^3}\right)^2 \frac{1}{c^5} a_2 \beta_2 I_{ij}^{(3)}(t)\int_{-\infty}^\infty \frac{d\tau}{\tau}[I_{ij}^{(4)}(t+\tau)-I_{ij}^{(4)}(t-\tau)]\nonumber\\
&=& \frac{G}{4}\left( \frac{G{\mathcal M}}{c^3}\right)^2 \frac{1}{c^5} \frac15 \frac{214}{105} I_{ij}^{(3)}(t)\int_{-\infty}^\infty \frac{d\tau}{\tau}[I_{ij}^{(4)}(t+\tau)-I_{ij}^{(4)}(t-\tau)]\,,\nonumber\\
H^{{\rm tail-of-tail}}_{J_2}(t)&=&-\frac{G}{4}\left( \frac{G{\mathcal M}}{c^3}\right)^2 \frac{1}{c^7}  b_2\beta_2 J_{ij}^{(3)}(t)\int_{-\infty}^\infty \frac{d\tau}{\tau}[J_{ij}^{(4)}(t+\tau)-J_{ij}^{(4)}(t-\tau)]\nonumber\\
&=&\frac{G}{4}\left( \frac{G{\mathcal M}}{c^3}\right)^2 \frac{1}{c^7}  \frac{16}{45}\frac{214}{105} J_{ij}^{(3)}(t)\int_{-\infty}^\infty \frac{d\tau}{\tau}[J_{ij}^{(4)}(t+\tau)-J_{ij}^{(4)}(t-\tau)]\,,\nonumber\\
H^{{\rm tail-of-tail}}_{I_3}(t)&=&-\frac{G}{4}\left( \frac{G{\mathcal M}}{c^3}\right)^2 \frac{1}{c^7}  a_3 \beta_3 I_{ijk}^{(4)}(t)\int_{-\infty}^\infty \frac{d\tau}{\tau}[I_{ijk}^{(5)}(t+\tau)-I_{ijk}^{(5)}(t-\tau)]
\nonumber\\
&=&\frac{G}{4}\left( \frac{G{\mathcal M}}{c^3}\right)^2 \frac{1}{c^7}  \frac{1}{189} \frac{26}{21} I_{ijk}^{(4)}(t)\int_{-\infty}^\infty \frac{d\tau}{\tau}[I_{ijk}^{(5)}(t+\tau)-I_{ijk}^{(5)}(t-\tau)]\,.
\eea
\end{widetext}
Introducing the  (Hilbert transform inspired) shorthand notation
\beq
H[X_l^{(n,m)}(t)] \equiv X_L^{(n)}(t)\int_{-\infty}^\infty \frac{dt'}{t'-t} X_L^{(m)}(t')\,,
\eeq
involving  principal-value integrals, the various contributions to the tail-of-tail Hamiltonian read
\bea
H^{{\rm tail-of-tail}}_{I_2}(t)
&=& \frac{G}{4}\left( \frac{G{\mathcal M}}{c^3}\right)^2 \frac{1}{c^5} \frac15 \frac{214}{105}   2H[I_2^{(3,4)}(t)] \,,\nonumber\\
H^{{\rm tail-of-tail}}_{J_2}(t)
&=&\frac{G}{4}\left( \frac{G{\mathcal M}}{c^3}\right)^2 \frac{1}{c^7}  \frac{16}{45}\frac{214}{105}  2 H[J_2^{(3,4)}(t)]\,,\nonumber\\
H^{{\rm tail-of-tail}}_{I_3}(t)
&=&\frac{G}{4}\left( \frac{G{\mathcal M}}{c^3}\right)^2 \frac{1}{c^7}  \frac{1}{189} \frac{26}{21}  2 H[I_3^{(4,5)}(t)]  \,.\nonumber\\
\eea
In the latter $t'$ integrals, one must replace $\frac{dt'}{t'-t}$ by its expression in terms of  $du'$, $u'$ and $u$
(expanding in powers of $e_t$).

Finally, we average over a period of the radial motion
\bea
\langle H^{{\rm tail-of-tail}}_{X_L}\rangle  \equiv \frac{\oint H^{{\rm tail-of-tail}}_{X_L}(t) dt }{\oint dt}\,,
\eea
with
\beq
\oint dt=2\pi  a_r^{3/2} +\sqrt{a_r}\pi\eta^2 (9-\nu)\,.
\eeq
We find
\bea
2\langle   H[I_2^{(3,4)}(t)] \rangle &=& \left(-\frac{4670}{3} e_t^2-128\right)\frac{\pi\nu^2}{a_r^{13/2}}\nonumber\\ 
&+& \left[\left(-\frac{158993}{21} e_t^2-\frac{4288}{7}\right)\nu \right.\nonumber\\
& +& \left. \frac{256805}{21} e_t^2+\frac{31168}{21} \right]\eta^2 \frac{\pi\nu^2}{a_r^{15/2}}\nonumber\\
2\langle   H[J_2^{(3,4)}(t)]\rangle  &=& -(1-4\nu)(30 e_t^2+1)  \frac{\pi\nu^2 }{a_r^{15/2}} \nonumber\\
2 \langle   H[I_3^{(4,5)}(t)]\rangle &=&  (1-4\nu)\left(\frac{910188}{5} e_t^2+\frac{49209}{5}\right)\frac{\pi\nu^2 }{a_r^{15/2}} \,. \nonumber \\
\eea

The final result then reads
\bea
\langle H^{{\rm tail-of-tail}}\rangle &=& \langle H^{{\rm tail-of-tail}}_{I_2}\rangle+\eta^2 \langle H^{{\rm tail-of-tail}}_{J_2}\rangle\nonumber\\
&+& \eta^2 \langle H^{{\rm tail-of-tail}}_{I_3}\rangle\nonumber\\
&=& 
\langle H^{{\rm tail-of-tail}}\rangle_{\eta^0}+\eta^2  \langle H^{{\rm tail-of-tail}}\rangle_{\eta^2}\,,\nonumber\\
\eea
where
\bea
\label{ris_fin_H}
\langle H^{{\rm tail-of-tail}}\rangle_{\eta^0}&=& \frac{\nu^2\pi}{a_r^{13/2}} \left(\frac{49969}{315} e_t^2    +\frac{6848}{525} \right)\nonumber\\ 
\langle H^{{\rm tail-of-tail}}\rangle_{\eta^2}&=& -\frac{\nu^2\pi}{a_r^{15/2}} \left[\left( \frac{1473251}{2450} e_t^2    +  \frac{4982}{315}\right)\nu\right.\nonumber\\  
&+&\left.  \frac{593849 }{630}e_t^2  +  \frac{991861}{7350} \right]\,. 
\eea
As usual, the transcription of this averaged tail-of-tail Hamiltonian into an equivalent EOB Hamiltonian
is obtained by equating
\bea
\langle H^{{\rm tail-of-tail}}_{\rm harmonic}\rangle = \frac{\oint H^{{\rm tail-of-tail}}(t) dt }{\oint dt}\,,
\eea
expressed as a function of $a_r$ and $e_t$ (or, as a function of energy and angular momentum), with its EOB counterpart
\bea
\label{Heobaver}
\langle H^{{\rm tail-of-tail}}_{\rm EOB}\rangle = \frac{\oint H^{{\rm tail-of-tail}}_{\rm EOB}(t) dt }{\oint dt}\,,
\eea
where $H^{{\rm tail-of-tail}}_{\rm EOB}$ is defined by its expression in terms of the EOB potentials 
$A(u,\nu)= A(u,\nu)^{\rm 1PN} + A(u,\nu)^{\rm tail-of-tail}$
and  ${\bar D}(u,\nu)= {\bar D}(u,\nu)^{\rm 1PN} + {\bar D}(u,\nu)^{\rm tail-of-tail}$.
The EOB potentials are then decomposed in powers of $u=\frac{GM}{R}$, according to
\bea
A(u,\nu) &=&1-2u + \sum_{n\geq 3} a_n(\nu, \ln u) u^n\,, \nonumber \\
{\bar D}(u,\nu)&=& 1 + \sum_{n\geq 2} {\bar d}_n(\nu, \ln u) u^n\,.
\eea
See Appendix \ref{EOB_Details} for details.

This leads to the following 5.5PN and 6.5PN tail-of-tail contributions to the EOB effective 
metric\footnote{We recall that $A_{6.5}$ belongs to the 5.5PN level, while ${\bar D}_{5.5}$
belongs to the 5.5PN level, etc.}
\bea
\label{ris_fins}
a_{6.5} &=&  \frac{13696}{525} \nu \pi \,,\nonumber\\
a_{7.5} &=& -  \frac{10052}{225}\nu^2\pi -\frac{512501}{3675}\nu \pi \,,\nonumber\\
{\bar d}_{5.5} &=&  \frac{264932}{1575}\nu \pi \,,\nonumber\\
{\bar d}_{6.5} &=& -\frac{893149}{2450}\nu^2\pi -\frac{21288791}{17640} \nu \pi\,. 
\eea
The right-hand sides of these results contain six numerical coefficients, four of them being proportional
to $\nu$ and two of them being proportional to $\nu^2$.
The four numerical coefficients entering the $O(\nu)$ contributions agree
with the 1SF results obtained in Refs. \cite{Bini:2014nfa,Bini:2016qtx}. [More precisely,
$a_{6.5}$ and the $\nu$ term in $a_{7.5}$ agree with Eqs. (11) and (14) in \cite{Bini:2014nfa}
(where a factor $\nu$ was factored out), while ${\bar d}_{6.5}$ and the $\nu$ term in ${\bar d}_{7.5}$ agree with Eq. (5) in \cite{Bini:2016qtx}.] These four numerical agreements are very satisfactory
and provide nice checks both of the structure of the tail-of-tail action \eqref{eqtailoftail0},
and of the value of the odd-parity beta coefficient $\beta_2^{(\rm odd)}$.
Note that, besides those checks, our work completes the previous 1SF results by providing
the full EOB 5.5PN and 6.5PN Hamiltonian (modulo $O(p_r^4)$ contributions) which includes
 2SF contributions in $a_{7.5}$ and ${\bar d}_{6.5}$.

\section{Tail-of-tail action and scattering angle at the 6.5PN accuracy}

We now consider hyperboliclike motions (i.e. scattering situations) and compute the tail-of-tail
contribution to the scattering angle $\chi$ (see Ref. \cite{Bini:2025vuk} for  state-of-the-art results
on the scattering angle). We work at the fractional 1PN accuracy, i.e. at the absolute
6.5 PN accuracy, and shall expand the scattering angle up to the 8th order in $G$.

As usual, it is convenient to express the scattering angle as a function of 
\beq
\label{pinfdef}
p_\infty \equiv \sqrt{\gamma^2-1} \equiv \frac{v}{\sqrt{1-v^2}}\,,
\eeq 
and
\beq
j \equiv \frac{J}{G m_1 m_2} = \frac{p_\infty \, b}{G M h}\,.
\eeq
Here $\gamma$ is the Lorentz factor,  $v$ is the relative velocity between the two incoming bodies, and
$h \equiv \frac{E}{M}= \sqrt{1+ 2 \nu (\gamma-1)}$. [Henceforth, we often use units such that $c=1$.]

We use the following notation for the PM expansion coefficients of the scattering angle:
\beq
\chi(p_\infty, j) = \sum_{n \geq 1} \frac{2 \chi_n(p_\infty)}{j^n}\,.
\eeq
We recall that the PN order of an nPM contribution $\propto \frac{p_\infty^m}{j^n}$ is $\frac12 (n+m)$.
For instance, the leading PN order (i.e., 5.5PN) contribution to the conservative scattering angle induced
by the leading tail-of-tail action (as computed in Ref. \cite{Bini:2020hmy}, see Eq. (5.10) there) reads (at the 5PM order)
\beq
\label{chi5p5}
\chi^{\rm LO}_{\rm tail-of-tail}=- \nu \frac{95872}{675}\frac{p_\infty^6}{j^5}+O(G^6)\,.
\eeq
The  tail-of-tail contribution to the (conservative) scattering angle is obtained
from the corresponding on-shell action contribution via
\beq \label{chivsS}
\chi^{{\rm tail-of-tail}}_{\rm time-split}=\frac{\partial}{\partial J}S^{{\rm tail-of-tail, on-shell}}_{\rm time-split}\,.
\eeq
Here, we generalize the leading-order result \eqref{chi5p5} of \cite{Bini:2020hmy} by going to the
6.5PN level and the 8PM one. 

To obtain  this result, we need to compute the fractionally 1PN-accurate
value of the tail-of-tail action \eqref{eqtailoftail3} along the 1PN-accurate hyperbolic motion.
It is convenient to use the frequency-domain version of \eqref{eqtailoftail3}. The  building blocks ${\sf G}_X^l$
read in frequency-space
\begin{widetext}
\bea
{\sf G}_X^l&=&\int_{-\infty}^\infty dt \int_{-\infty}^\infty \frac{d\tau}{\tau}X_L^{(l+1)}(t)[X_L^{(l+2)}(t+\tau)-X_L^{(l+2)}(t-\tau)]\nonumber\\
&=& \int_{-\infty}^\infty \frac{d\tau}{\tau} \int \frac{d\omega}{2\pi}(-i\omega)^{l+1}\hat X_L(\omega)[e^{i\omega   \tau }(i\omega)^{l+2}\hat X_L(-\omega)
-e^{-i\omega   \tau }(i\omega)^{l+2}\hat X_L(-\omega)]\nonumber\\
&=& 4i (-1)^{l+1} i^{2l+3}\int \frac{d\omega}{2\pi} \omega^{2l+3}|\hat X_L(\omega)|^2  \int_{0}^\infty \frac{d\tau}{\tau}\sin (\omega   \tau)\nonumber\\
&=& 4 (-1)^{l+1} (-1)^l\int \frac{d\omega}{2\pi} \omega^{2l+3}|\hat X_L(\omega)|^2  \frac{\pi}{2}{\rm sgn}(\omega)\,,
\eea
\end{widetext}
where we used
\beq
\int_{0}^\infty \frac{d\tau}{\tau}\sin (\omega   \tau)=\frac{\pi}{2}{\rm sgn}(\omega)\,.
\eeq
This yields
\bea
{\sf G}_X^l&=& 
 -2\int_0^\infty  d\omega   \omega^{2l+3}|\hat X_L(\omega)|^2  \,.
\eea
Inserting this result in \eqref{eqtailoftail3} we find the (fractional) 1PN level expressions
\begin{widetext}
\bea
\label{eqtailoftail4}
S^{{\rm tail-of-tail}}_{\rm time-split}&=& \frac{G}4  \left(\frac{G{\mathcal M}}{c^3}\right)^2 \frac{1}{c^5}\left[ 
\left( -\frac15 \frac{214}{105} {\sf G}_I^2
-\frac{1}{c^2}\frac{16}{45}\, \frac{214}{105}  {\sf G}_J^2\right)-
\frac{1}{c^2}\frac{1}{189}\, \frac{26}{21} {\sf G}_I^3\right]+\ldots\nonumber\\
&=& \frac{G}4  \left(\frac{G{\mathcal M}}{c^3}\right)^2 \frac{1}{c^5}\left[ 
\left( -\frac15 \frac{214}{105}    (-2)\int_0^\infty  d\omega   \omega^{7}|\hat I_{ij}(\omega)|^2  
-\frac{1}{c^2}\frac{16}{45}\, \frac{214}{105} (-2)\int_0^\infty  d\omega   \omega^{7}|\hat J_{ij}(\omega)|^2  \right)\right.\nonumber\\
&-&\left.
\frac{1}{c^2}\frac{1}{189}\, \frac{26}{21} (-2)\int_0^\infty  d\omega   \omega^{9}|\hat I_{ijk}(\omega)|^2
\right]\,,
\eea
\end{widetext}
that is
\bea
\label{eqtailoftail5}
S^{{\rm tail-of-tail}}_{\rm time-split}
&=& G \left(\frac{G{\mathcal M}}{c^3}\right)^2 \frac{1}{c^5} \int_0^\infty  d\omega {\mathcal F}_{\rm tail-of-tail}(\omega)\,, \nonumber \\
\eea
where ${\mathcal M} = M + M\nu p_\infty^2\eta^2/2+O(\eta^4)$ and where
we defined
\bea
{\mathcal F}_{\rm tail-of-tail}(\omega)&\equiv& 
   \frac15 \frac{107}{105}     \omega^{7}|\hat I_{ij}(\omega)|^2  \nonumber\\
&+& \frac{1}{c^2}\left(\frac{16}{45}\, \frac{107}{105}   \omega^{7}|\hat J_{ij}(\omega)|^2 \right.\nonumber\\   
&+&\left. \frac{1}{189}\, \frac{13}{21}  \omega^{9}|\hat I_{ijk}(\omega)|^2\right)\,. \nonumber \\
\eea
The result \eqref{eqtailoftail5} is very similar
to the frequency-domain expression of the 
fractional 1PN expression of the linear-tail contribution
to the radiated energy, $E^{\rm rad}_{\rm tail}$, which differs from  \eqref{eqtailoftail5}  in having only
the first power of $\frac{G{\mathcal M}}{c^3}$ in prefactor, and a slightly different 
integrand ${\mathcal F}(\omega)$, namely (see, e.g., Appendix D of \cite{Bini:2021gat} and
Ref. \cite{Bini:2022xpp})
\bea
{\mathcal F}_{\rm tail}(\omega)&=& 2\left[ 
   \frac15     \omega^{7}|\hat I_{ij}(\omega)|^2   \right.\nonumber\\
&+&\left.\frac{1}{c^2}\left(\frac{16}{45}\, \omega^{7}|\hat J_{ij}(\omega)|^2
+\frac{1}{189}\,   \omega^{9}|\hat I_{ijk}(\omega)|^2\right)\right]\,.\nonumber\\
\eea
[The similarity comes from the fact that
the (time-symmetric projection of the) tail kernel is $|\omega|$; see Appendix D of  \cite{Bini:2021gat}.]
It is then easy to  modify the computations of  $E^{\rm rad}_{\rm tail}$ in 
Refs. \cite{Bini:2021gat,Bini:2022xpp}  to derive the on-shell value of the tail-of-tail action.

We recall that the latter computation uses the quasi-Keplerian representation of the 1PN-accurate
hyperbolic motion (obtained from Eq. \eqref{quasiKeplerian} by analytic continuation: $u \to i \, v $).
The frequency-domain 1PN-accurate multipole moments are then expressed in terms of modified Bessel functions $K_{i \frac{u}{e_t}}(u)$
and $K_{1+ i \frac{u}{e_r}}(u)$, where $u$ now denotes the frequency variable 
\beq
 u\equiv \omega e_r \bar a_r^{3/2}\,.
\eeq
Taking the large-eccentricity expansion of the integrand finally generates Bessel functions $K_0(u)$ and $K_1(u)$  at the leading order, and their derivatives with respect to the order at higher orders in $\frac1{e_r}$. 
From a practical point of view, the integration over the frequencies is more easily performed by using a Mellin transform.
Our final result at the fractional 1PN accuracy is the following
\begin{widetext}
\bea
S^{{\rm tail-of-tail, on-shell}}_{\rm time-split}
&=& G M^2 \nu^2\left\{
\left[\left(-\frac{1696 \nu }{175}-\frac{377824}{33075}\right)
   p_\infty^8+\frac{23968 p_\infty^6}{675}\right]\frac{1}{j^4}\right.\nonumber\\
&+&\left[\left(-\frac{37017 \pi ^2 \nu }{9800}-\frac{1288611 \pi
   ^2}{78400}+\frac{164352}{1225}\right) p_\infty^7+\frac{10593 \pi ^2
   p_\infty^5}{1400}\right]\frac{\pi}{j^5}\nonumber\\
&+&\left[\left(\left(-\frac{10432}{189}-\frac{9673472 \pi ^2}{165375}\right) \nu +\frac{7383 \zeta (3)}{25}-\frac{771584
   \pi
   ^2}{14175}+\frac{1680022}{11025}\right)p_\infty^6\right.\nonumber\\
&+&\left.
\left(\frac{499904}{47
   25}+\frac{4738816 \pi ^2}{70875}\right)p_\infty^4\right]\frac{1}{j^6} \nonumber\\
&+&
\left[\left(\left(\frac{384117 \pi ^4}{12544}-\frac{724127 \pi
   ^2}{1960}\right) \nu +\frac{18385421 \pi
   ^4}{150528}-\frac{1874672707 \pi
   ^2}{1552320}+\frac{11244416}{23625}\right)
   p_\infty^5\right.\nonumber\\
&+&\left.\left.
\left(\frac{168953 \pi ^2}{630}-\frac{58957 \pi
   ^4}{2688}\right) p_\infty^3\right]\frac{\pi}{j^7}
+O\left(\frac{1}{j^8}\right)
\right\}\,.
\eea
Using Eq. \eqref{chivsS}, the corresponding tail-of-tail scattering angle reads
\bea  \label{chitailoftail}
\chi^{{\rm tail-of-tail}}_{\rm time-split}&=&\nu \left\{  \left[ \left( \frac{6784\nu}{175} + \frac{1511296}{33075}\right)p_\infty^8 - \frac{95872}{675}p_\infty^6  \right]\frac{1}{j^5} \right.\nonumber\\
&+& \pi \left[\left(\frac{37017\nu\pi^2}{1960} - \frac{164352}{245}  + \frac{1288611}{15680}\pi^2\right)p_\infty^7 -  \frac{10593p_\infty^5 \pi^2}{280} \right]\frac{1}{j^6} \nonumber\\
&+& \left[\left(\frac{20864\nu}{63} + \frac{19346944\pi^2\nu}{55125} - \frac{3360044}{3675} + \frac{1543168\pi^2}{4725} - \frac{44298\zeta(3)}{25}\right)p_\infty^6
\right.\nonumber\\ 
&+&\left. \left(-\frac{999808}{1575} - \frac{9477632\pi^2}{23625}\right)p_\infty^4\right]\frac{1}{j^7}\nonumber\\ 
&+& \pi \left[\left(-\frac{384117}{1792}\nu\pi^4 + \frac{724127}{280}\nu\pi^2 - \frac{18385421}{21504}\pi^4 - \frac{11244416}{3375} + \frac{1874672707}{221760}\pi^2\right)p_\infty^5\right.\nonumber\\ 
&+&\left.\left. \left(\frac{58957}{384}\pi^4 - \frac{168953}{90}\pi^2\right)p_\infty^3\right]\frac{1}{j^8}+O\left(\frac{1}{j^9}\right)
\right\}\,.
\eea
This result extends the  previously known leading-order term, Eq. \eqref{chi5p5}, 
both by including fractional 1PN corrections (thereby reaching the absolute 6.5PN accuracy),
 and higher powers of $\frac1j$, i.e., higher powers of $G$, up to the $G^8$ level included.

\section{Comparison with the 5PM-1SF results of Ref. \cite{Driesse:2024feo}}

In the previous Section we computed the contribution of the time-symmetric tail-of-tail action to the
 scattering angle. This contribution fully describes the 5.5PN-plus-6.5PN contributions (at
 PM orders $G^5$, $G^6$, $G^7$ and $G^8$) to the  {\it conservative} scattering angle. 
 This result thereby extends the previous 6PN-accurate computation of the scattering angle
 obtained within the Tutti Frutti approach  \cite{Bini:2020wpo,Bini:2020hmy} by half a PN order.
 In particular, using the standard notation (which factors  out a factor 2 in the definition of 
 the nPM coefficient $\chi_n$)
 \beq
\chi(p_\infty, j) = \sum_{n \geq 1} \frac{2 \chi_n(p_\infty)}{j^n}\,,
\eeq
we now know the {\it conservative} part of the $G^5$ coefficient $\chi_5$ to the absolute 6.5PN accuracy, namely
\bea \label{chi5consTF}
\chi_5^{\rm cons, TF}(p_\infty)&=&  \frac{1}{5 p_\infty^5}-\frac{2}{p_\infty^3}+\frac{ 32-8\nu }{p_\infty}\nonumber\\
&+&\left[320+24\nu^2+\left(-\frac{1168}{3}+\frac{41}{8}\pi^2\right)\nu  \right] p_\infty\nonumber\\
&+&\left[640-40\nu^3+\left(\frac{7342}{9}-\frac{287}{24}\pi^2\right)\nu^2
+\left(-\frac{6272}{45} \ln(2 p_\infty)-\frac{247027}{135}+\frac{5069}{144}\pi^2\right)\nu \right]  p_\infty^3\nonumber\\
&+&\left[\frac{1792}{5}+56\nu^4+\left(-\frac{11108}{9}+\frac{451}{24}\pi^2 \right)\nu^3
+\left(\frac{325838}{105}-\frac{40817}{640}\pi^2+\frac{13952}{45} \ln(2 p_\infty)-\frac{4}{15} \bar d_5^{\nu^2}\right)\nu^2\right.\nonumber\\
&+&\left. \left(-\frac{21106609}{7875}+\frac{111049}{960}\pi^2-\frac{74432}{525} \ln(2 p_\infty) \right) \nu\right] p_\infty^5\nonumber\\
&-&\frac{\nu}{2} \frac{95872}{675}p_\infty^6\nonumber\\
&+&\left[-72\nu^5+\left(-\frac{205}{8}\pi^2+\frac{4934}{3}\right)\nu^4
+\left(-\frac{1575083}{378}+\frac{256133}{2880}\pi^2+\frac{8}{15}\bar d_5^{\nu^2} -\frac{21632}{45} \ln(2 p_\infty) \right)\nu^3\right.\nonumber\\
&+& \left(\frac{288224}{1575} \ln(2 p_\infty)+\frac{408095609}{132300}-\frac{69239}{448}\pi^2-\frac{4}{35} q_{45}^{\nu^2} \right)\nu^2\nonumber\\
&+&\left. \left(-\frac{881392}{11025} \ln(2 p_\infty)+\frac{689996443}{6174000}-\frac{184881}{4480}\pi^2 \right)\nu\right] p_\infty^7
\nonumber\\
&+&\nu \left( \frac{6784\nu}{175} + \frac{1511296}{33075}\right)p_\infty^8 
+O(p_\infty^9)
\,.
\eea
  Recently, Ref. \cite{Driesse:2024feo} computed
 the (relative) scattering angle at the PM order $G^5$,  at all orders in the PN expansion, but only
 at the first self-force (1SF) order. However, the result of Ref. \cite{Driesse:2024feo} includes both
 conservative and 1SF radiative effects. Some time ago Ref. \cite{Bini:2012ji} showed that, when treating
dynamical radiation-reaction effects to linear order (as is the case in a 1SF treatment), there exists a
simple universal link between the radiative losses of energy and angular momentum and the radiation-reaction
contribution to the scattering angle, namely
\beq \label{chirrlin}
\chi^{\rm rr, lin} = - \frac12 \left( E^{\rm rad} \frac{\partial \chi^{\rm cons}(E,J)}{\partial E}  +   J^{\rm rad} \frac{\partial \chi^{\rm cons}(E,J)}{\partial J} \right)\,.
\eeq

In order to compare our new, conservative 6.5PN-accurate 5PM result \eqref{chi5consTF}
with the  radiation-reacted 5PM-1SF results of Ref. \cite{Driesse:2024feo}, we need to add to our
result the (linear) radiation-reaction contribution \eqref{chirrlin}.

In the 1SF limit the 5PM coefficient of $\chi^{\rm rr, lin}$, Eq.  \eqref{chirrlin}, reads
\bea
\chi_5^{\rm rr, lin, 1SF}&=&\nu\left[-\frac{\gamma  \left(2 \gamma ^2-3\right) \hat E_4^0}{2\left(\gamma ^2-1\right)^{3/2}}
+\frac{3}{8} \pi  \left(5\gamma ^2-1\right) \hat J_3
+\frac{\left(2 \gamma^2-1\right) \hat J_4^0}{2 \sqrt{\gamma ^2-1}}
-\frac{15}{8}\pi ^2 \gamma  \left(\gamma ^2-1\right)^{3/2}\mathcal E\right.\nonumber\\
&+&\left.
\frac{\left(64 \gamma ^6-120 \gamma ^4+60\gamma ^2-5\right) \hat J_2}{2 \left(\gamma^2-1\right)^{3/2}}\right]\,.
\eea
Here $E_n$ and $J_n$ denote the (dimensionless) PM expansion coefficients of 
$ E^{\rm rad}$ and $ J^{\rm rad}$ defined according to
\bea
 \frac{E^{\rm rad}}{M} &=& \nu^2 \sum_{n=3}^\infty \frac{E_n}{j^n}\,, \nonumber \\
  \frac{J^{\rm rad}}{J} &=&  \nu \sum_{n=2}^\infty \frac{J_n}{j^n} \,,
\eea
and we have used  the $\nu$-structure of the relevant $E_n$ and $J_n$, namely 
\bea
h^4E_3&=&\pi p_\infty^3\mathcal E(\gamma)
\,,\nonumber\\
h^5E_4&=&\hat E_4^0(\gamma)+\nu \hat E_4^1(\gamma)
\,,\nonumber\\
h^2J_2&=&\hat J_2(\gamma)
\,,\nonumber\\
h^3J_3+\nu h^2E_3&=&\hat J_3(\gamma)
\,,\nonumber\\
h^4J_4+\nu h^3E_4&=&\hat J_4^0(\gamma)+\nu \hat J_4^1(\gamma)
\,.
\eea
Among these coefficients, the 3PM and 4PM energy coefficients are known exactly (see Refs. \cite{Herrmann:2021tct} and \cite{Dlapa:2022lmu}, respectively)  as well as the 2PM and 3PM angular momentum coefficients (see Refs. \cite{Damour:2020tta}) and \cite{Manohar:2022dea}, respectively).
By contrast, the 4PM angular-momentum loss coefficient  $J_4$ has not yet been fully derived.
Recently,  the  part of $J_4(p_\infty)$ involving even powers of $p_\infty$ (starting at $O(p_\infty^4)$) was
derived in Ref. \cite{Heissenberg:2025ocy}. Very recently, the 1SF contribution 
of the remaining part of 
$J_4(p_\infty)$ (involving odd powers of $p_\infty$) has been computed up to $p_\infty^{15}$
in Ref. \cite{Geralico:2025rof} by using black-hole perturbation techniques (the previous PN knowledge
of $J_4(p_\infty)$  was limited to the $p_\infty^{7}$ level \cite{Bini:2022enm}).

Combining these results, we find
\bea
\chi_5^{\rm rr, lin, 1SF}&=&\nu\left[
\frac{1504}{45}
+\left(\frac{43184}{525}+\frac{42 \pi ^2}{5}\right) p_\infty^2
+\frac{3584}{45} p_\infty^3
+\left(\frac{1741664}{11025}+\frac{267 \pi^2}{14}\right) p_\infty^4
+\frac{2496}{35} p_\infty^5\right.\nonumber\\
&&\left.
+\left(\frac{541568}{3465}+\frac{51847 \pi^2}{6720}\right) p_\infty^6
+\frac{37664}{1575} p_\infty^7
+\left(\frac{94791296}{3468465}-\frac{1786689 \pi^2}{197120}\right) p_\infty^8
+O(p_\infty^9)\right]+ O(\nu^2).
\eea

\end{widetext}

We then compared the so obtained 6.5PN-accurate knowledge of the 1SF radiation-reacted
scattering angle
\beq
\chi_5(p_\infty) = \chi_5^{\rm cons, TF}(p_\infty)+  \chi_5^{\rm rr, lin, 1SF}(p_\infty) + O(\nu^2),
\eeq
with the result obtained  in Ref. \cite{Driesse:2024feo}, and written there in the form
\beq
\chi^{\rm (5PM)}= \left(\frac{GM}{c^2 b} \right)^5 h [\chi^{(5,0)}+\nu \chi^{(5,1)}+\nu^2 \chi^{(5,2)}+\frac{\nu^3}{h^2}\chi^{(5,3)}]\,,
\eeq
where the 1SF correction $\chi^{(5,1)}$ is given in Eq. (S.30a) there. [We recall
that $v=\frac{p_\infty}{\sqrt{1+p_\infty^2}}$ and  that $\frac{GM}{c^2 b}=\frac{p_\infty}{hj}$.]

We find complete agreement up to the 6.5PN order. Of particular significance is
the agreement of the last term at the 6.5PN level which is new with this work.
In the recent work \cite{Bini:2025vuk} we had checked the agreement up to the
6PN level included. The extra half-PN-order check up to the 6.5PN level is important
because it involves both the fractionally 1PN contribution of the tail-of-tail action \eqref{eqtailoftail0},
and the very recent high-PN computation of $J_4$ \cite{Geralico:2025rof}.

\section{Concluding remarks}

We have derived several physical quantities at
the 6.5 PN level from the recently obtained all-multipole tail-of-tail action: 
Delaunay Hamiltonian for slightly eccentric bound orbits (together
with its EOB transcription), and half-integer PN contributions to
the conservative scattering angle. We have then provided several checks of our results against
previous results. 

The $O(\nu)$ part of the 5.5PN and 6.5PN contributions to the EOB Hamiltonian 
(modulo $O(p_r^4)$) were found to agree with previous 1SF results obtained by black-hole perturbation
methods (see Section \ref{secelliptic}). [These checks give in passing a check of the value of the
odd-parity quadrupolar beta coefficient.]
Beyond these checks, our  results provide a more complete knowledge of
the EOB Hamiltonian, including  $O(\nu^2)$ contributions (see Eqs. \eqref{ris_fins}).

Concerning unbound motions, we have found agreement at the 6.5PN level
with the 1SF 5PM result of Ref. \cite{Driesse:2024feo} when completing our tail-of-tail-derived
conservative scattering angle with the linear-response radiation-reaction contribution.
Importantly, the 6.5PN knowledge of the latter was obtained by using the very recent
result of Ref. \cite{Geralico:2025rof}.

  Our scattering results go well beyond the $G^5$ level to reach the $G^8$ level.
  We note in passing the interesting appearance of $\zeta(3)$ at the $G^7$ and 6.5PN  levels
  (see Eq. \eqref{chitailoftail}). It would be interesting to check this result against other computations.

\appendix

\section{Multipolar moments: PN-expanded and eccentricity-expanded expressions}
\label{mult_mom}

Let us consider  the quantities
\bea
I_2^{(3,4)}(u,u')&=&I_{ij}^{(3)}(t)I_{ij}^{(4)}(t')\big|_{t=t(u),t'=t'(u')}\,, \nonumber\\
J_2^{(3,4)}(u,u')&=& J_{ij}^{(3)}(t)J_{ij}^{(4)}(t')\big|_{t=t(u),t'=t'(u')}\,, \nonumber\\
I_3^{(4,5)}(u,u')&=& I_{ijk}^{(4)}(t)I_{ijk}^{(5)}(t')\big|_{t=t(u),t'=t'(u')}\,,
\eea
PN-expanded (up to the fractional 1PN level) and small-eccentricity-expanded (up to $O(e_t^2)$).
Each such quantity, say $X(u,u')$, can be written as (denoting $\eta \equiv \frac1c$)
\bea
X(u,u')&=& [X_{(\eta^0,e_t^0)}(u,u')+e_t X_{(\eta^0,e_t^1)}(u,u')\nonumber\\
&+& e_t^2 X_{(\eta^0,e_t^2)}(u,u')]\nonumber\\
&+& [X_{(\eta^2,e_t^0)}(u,u')+e_t X_{(\eta^2,e_t^1)}(u,u')\nonumber\\
&+& e_t^2 X_{(\eta^2,e_t^2)}(u,u')]\eta^2\,,
\eea
which involves six coefficients: $X_{(\eta^0,e_t^0)}(u,u')$, $X_{(\eta^0,e_t^1)}(u,u')$, 
$X_{(\eta^0,e_t^2)}(u,u')$, $X_{(\eta^2,e_t^0)}(u,u')$, $X_{(\eta^2,e_t^1)}(u,u')$, $X_{(\eta^2,e_t^2)}(u,u')$.
The results are summarized in the following Table \ref{tab:1}. Clearly, for multipolar moment contributions which are already at 1PN the needed expansion is truncated at the Newtonian level: this is the case of $J_2$ and $I_3$, whereas the knowledge of $I_2$ is needed trough the 1PN level of accuracy.

\begin{table*}  
\caption{\label{tab:1}  Multipolar-moment bilinears entering the split tail-of-tail action, PN-expanded (up to 1PN) and eccentricity-expanded (up to $O(e_t^2)$). They are evaluated in harmonic coordinates  and by using the 1PN quasi-Keplerian parametrization of  elliptic-like orbits.}
\begin{ruledtabular}
\begin{tabular}{l|l}
$I_{2(\eta^0,e_t^0)}^{(3,4)}(u,u') $&$\frac{64\nu^2}{ a_r^{13/2}}\sin(2 (u - u'))$\\ 
$I_{2(\eta^0,e_t^1)}^{(3,4)}(u,u')$ &$\frac{\nu^2}{a_r^{13/2}}\left[40\sin(u-2 u')+260\sin(2 u-3 u')+52\sin(2 u-u')+152\sin(3 u-2 u')\right]$\\ 
$I_{2(\eta^0,e_t^2)}^{(3,4)}(u,u')$ &$\frac{\nu^2}{a_r^{13/2}}\left[ 26\sin(2 u)-20\sin(2 u')+\frac{325}{2}\sin(u-3 u')+\frac{197}{6}\sin(u-u')+\frac13 \sin(u+u')
+570\sin(2 u-4 u')\right.$\\
&$\left.+380\sin(2 u-2 u')+\frac{1235}{2}\sin(3 u-3 u')+220\sin(4 u-2 u')+\frac{247}{2}\sin(-u'+3 u)\right]$\\ 
$I_{2(\eta^2,e_t^0)}^{(3,4)}(u,u')$ &$\frac{\nu^2}{ a_r^{15/2}}\frac{32}{21}\left[ 252(u -u')\cos(2(u - u')) +  (-487 + 201\nu)\sin(2(u-u'))   \right]$\\ 
$I_{2(\eta^2,e_t^1)}^{(3,4)}(u,u')$ &$\frac{\nu^2}{ a_r^{15/2}} \left[\frac{2}{21} 2520( u- u')\cos(u-2 u')+\frac{2}{21} 16380(- u'+u)\cos(2 u-3 u')
+\frac{2}{21} 3276( u- u')\cos(2 u-u')\right.$\\
&$+\frac{2}{21} 9576 (u- u')\cos(3 u-2 u')+\frac{2}{21} (-6334+2286\nu)\sin(u-2 u')
+\frac{2}{21} (-26705+12579\nu)\sin(2 u-3 u')$\\
&$\left.+\frac{2}{21} (2751\nu-6433)\sin(2 u-u')+\frac{2}{21} (-13610+7314\nu)\sin(3 u-2 u')\right]$\\
$I_{2(\eta^2,e_t^2)}^{(3,4)}(u,u')$ &$\frac{\nu^2}{ a_r^{15/2}} \left[ \frac{1}{84} 13104( u-u')\cos(2 u)+\frac{1}{84}10080(- u'+ u)\cos(2 u')
+\frac{1}{84}81900( u- u')\cos(u-3 u')\right.$\\
&$+\frac{1}{84} 16380(- u'+ u)\cos(u-u')+\frac{1}{84} 287280( u- u')\cos(2 u-4 u')
+\frac{1}{84}223776(- u'+  u)\cos(2 u-2 u')$\\
&$ +\frac{1}{84}311220(- u'+ u)\cos(3 u-3 u')+\frac{1}{84}110880 (- u'+  u)\cos(4 u-2 u')
+\frac{1}{84}62244 (- u'+  u)\cos(-u'+3 u)$\\
&$ +\frac{1}{84} (-20692+11004\nu)\sin(2 u)+\frac{1}{84} (23320-9144\nu)\sin(2 u')
+\frac{1}{84} (-181105+71865\nu)\sin(u-3 u')$\\
&$+\frac{1}{84} (-42359+15595\nu)\sin(u-u')+\frac{1}{84} (-678+46\nu)\sin(u+u')+\frac{1}{84} (210396\nu-373540)\sin(2 u-4 u')$\\
&$ +\frac{1}{84} (-335304+173640\nu)\sin(2 u-2 u')+\frac{1}{84} (228471\nu-348275)\sin(3 u-3 u')+\frac{1}{84}(80040\nu-105640)\sin(4 u-2 u')$\\
&$\left.+\frac{1}{84} (50163\nu-90403)\sin(-u'+3 u)\right]$\\ 
\hline
$J_{2(\eta^0,e_t^0)}^{(3,4)}(u,u') $&$\frac{\nu^2 (1 - 4\nu)}{2 a_r^{15/2}}\sin(u - u')$\\  
$J_{2(\eta^0,e_t^1)}^{(3,4)}(u,u')$ &$-\frac{\nu^2}{ a_r^{15/2}}(1-4\nu) \left[ -\frac14 \sin(u)+\frac14 \sin(u')-\frac{15}{4} \sin(u-2 u')
-\frac74 \sin(2 u-u') \right]$\\ 
$J_{2(\eta^0,e_t^2)}^{(3,4)}(u,u')$ &$-\frac{\nu^2}{ a_r^{15/2}}(1-4\nu) \left[ -\frac78   \sin(2 u)+\frac{15}{8} \sin(2 u')
-\frac{45}{4} \sin(u-3 u')-\frac{19}{4}  \sin(u-u')-\frac18   \sin(u+u')\right.$\\
&$\left.-\frac{105}{8}  \sin(2 u-2 u')-\frac{25}{8}  \sin(-u'+3u) \right]$\\ 
\hline
$I_{3(\eta^0,e_t^0)}^{(3,4)}(u,u') $&$\frac{3\nu^2 ( 1 -4\nu)}{20 a_r^{15/2}} \left[32805\sin(3(u - u')) + \sin( u-u') \right]$\\  
$I_{3(\eta^0,e_t^1)}^{(3,4)}(u,u')$ &$-\frac{\nu^2}{ a_r^{15/2}}(1-4\nu) \left[-\frac{3}{40} \sin(u)+\frac{3}{40}  \sin(u')+\frac{99}{40}  \sin(u-2 u')
-\frac{24057}{8}  \sin(2 u-3 u')+\frac{51}{40}  \sin(2 u-u')\right. $\\
&$\left. -\frac{35721}{8}  \sin(3 u-2 u')-\frac{127575}{8} \sin(4 u-3 u')
-\frac{189783}{8 } \sin(-4 u'+3 u)\right]$\\ 
$I_{3(\eta^0,e_t^2)}^{(3,4)}(u,u')$ &$-\frac{\nu^2}{ a_r^{15/2}}(1-4\nu) \left[ \frac{51}{80} \sin(2 u)-\frac{99}{80} \sin(2 u')
-\frac{25335}{16}  \sin(u-3 u')+\frac{123}{40}  \sin(u-u')+\frac{33}{80 } \sin(u+u')\right. $\\
&$ -\frac{231957}{16}  \sin(2 u-4 u')
-\frac{109989}{40}  \sin(2 u-2 u')-\frac{991683}{16}  \sin(3 u-5 u')-\frac{279693}{8} \sin(3 u-3 u')-\frac{231525}{16}  \sin(4 u-2 u')$\\
&$\left. -\frac{120285}{4}  \sin(5 u-3 u')-\frac{1230075}{16}  \sin(-4 u'+4 u)-\frac{220863}{80}  \sin(-u'+3 u)\right]$\\ 
 \end{tabular}
\end{ruledtabular}
\end{table*}

\section{Details on the Delaunay averaging of the EOB Hamiltonian}
\label{EOB_Details}

The EOB (real) Hamiltonian, $H$, reads
\beq
H=Mc^2 \sqrt{1+2\nu (\hat H_{\rm eff}-1)}\,,
\eeq
in terms of the effective Hamiltonian $H_{\rm eff}=\mu \hat H_{\rm eff}$. 
Here $M=m_1+m_2$ is the total mass of the system, $\mu=m_1m_2/(m_1+m_2)$ is the reduced mass and $\nu=\mu/M$ the symmetric mass ratio.

As standard, the (rescaled) phase-space variables (for equatorial motions) are $(r,\phi)$ or $(u=1/r,\phi)$ and $(p_r,p_\phi)$.
In the so-called DJS gauge or $p_r-$gauge, the azimuthal momentum  $p_\phi \equiv j$ enters 
$H_{\rm eff}^2/A(u,\nu)$ only through the centrifugal term $\frac{p_\phi^2}{r^2}= j^2 u^2$. Moreover, the effective Hamiltonian introduces several potentials, $A(u)$, $\bar D(u)$, and $Q(u,p_r)$:
\beq
\hat H_{\rm eff}=\sqrt{A(u) (1+j^2u^2+p_r^2A(u) \bar D(u)+Q(u,p_r))}\,,
\eeq
where $A(u)$ is the \lq\lq main" radial EOB potential
\bea
A(u)&=&1-2u +\nu a^{\nu^1}(u)+\nu^2 a^{\nu^2}(u)+\nu^3 a^{\nu^3}(u)+\ldots\,,
\eea
and $\bar D(u)$ is the first potential associated with non-radial motion
\bea
\bar D(u)&=&1+\nu \bar d^{\nu^1}(u)+\nu^2 \bar d^{\nu^2}(u)+\nu^3 \bar d^{\nu^2}(u)+\ldots\,.
\eea
Finally,
\beq
Q(u, p_r)=p_r^4 q_4(u)+p_r^6 q_6(u)+p_r^8 q_8(u)+p_r^{10}q_{10}(u)+\ldots.
\eeq
where
\begin{eqnarray}
q_{2n}(u)&=& \nu q_{2n}^{\nu^1}(u)+\nu^2 q_{2n}^{\nu^2}(u)+\nu^3 q_{2n}^{\nu^3}(u)+\ldots\,,
\end{eqnarray}
with $n=2,3,..$.
For the purpose of the present 1PN and $O(e_t^2)$ ($\sim O(p_r^2)$) analysis, 
the potential $Q = O(p_r^4)$ is not needed.

Hence, the effective Hamiltonian simplifies to
\beq
\hat H_{\rm eff}=\sqrt{A(u) (1+j^2u^2+p_r^2A(u) \bar D(u))}\,,
\eeq
and can be rewritten as the sum of an unperturbed Hamiltonian 
\beq
\hat H_{\rm eff}^{(0)}=\sqrt{(1-2u)(1+j^2u^2+(1-2u)p_r^2 )}\,,
\eeq
(which uses only 1PN information for the potentials $A=1-2u+O(u^3)$ and $\bar D=1+O(u^2)$), and a first-order (tail-of-tail, say order $\epsilon$) perturbation, $\hat H_{\rm eff}^{(1)}$, which implies a correction to both these potentials as
\beq
A\to A=1-2u+\epsilon \delta A\,,\qquad \bar D \to \bar D=1+\delta \bar D\,,
\eeq
with
\bea
\delta A&=& a_{6.5} u^{13/2}+a_{7.5} u^{15/2}\,,\nonumber\\
\delta \bar D &=& {\bar d}_{5.5} u^{11/2}+{\bar d}_{6.5} u^{13/2}\,.
\eea

To first order in $\epsilon$ the real Hamiltonian then reads $H=H^{(0)}+\epsilon H^{(1)}$, with
\bea
\frac{H^{(0)}}{\nu}&=&\frac12 (p_r^2+p_\phi^2 u^2 -2u)+\eta^2 \left[
-\frac{\nu+1}{8} u^4 p_\phi^4\right.\nonumber\\
&+&\frac12 (\nu-1)p_\phi^2 u^3-\frac{\nu+1}{2}\left(1+\frac12 p_\phi^2 p_r^2\right)u^2 \nonumber\\
&+&\left. \frac{\nu-3}{2}u p_r^2-\frac{\nu+1}{8}p_r^4\right]\,,
\eea
and
\bea
\label{H1eob}
\frac{H^{(1)}}{\nu}&=& \frac12  (p_r^2 {\bar d}_{5.5}+a_{6.5} u)u^{11/2}\nonumber\\
&+&\eta^2  \left[
-\frac14  (\nu +1){\bar d}_{5.5}  p_r^4 u^{11/2}\right.\nonumber\\
&+&\left(\frac12 (\nu-3){\bar d}_{5.5}-\frac14 (\nu-3)a_{6.5}+\frac12 {\bar d}_{6.5}\right)p_r^2 u^{13/2}\nonumber\\
&+& \left(-\frac{ \nu +1}{4} {\bar d}_{5.5} p_r^2 p_\phi^2
+\frac{\nu+1}2   a_{6.5}+\frac12   a_{7.5}\right) u^{15/2}\nonumber\\
&+&\left.\frac14  (1 -\nu)a_{6.5} p_\phi^2 u^{17/2}\right]\,.
\eea

The unperturbed motion can be still written in the quasi-Keplerian form \eqref{quasiKeplerian}, but with orbital parameters now referring to EOB coordinates (hence labelled with a superscript $e$).
The 1PN conserved energy $H^{(0)}=E$ and angular momentum $p_\phi^{(0)}=L$ read
\bea
E&=& -\frac{1}{2 a_r^e}+\eta^2 \frac{3-\nu}8 \frac{1}{a_r^e{}^2}\,,\nonumber\\
L &=& \sqrt{a_r^e(1-e_r^e{}^2)} +\eta^2\frac{\frac32 +\frac12 e_r^e{}^2 }{\sqrt{a_r^e(1-e_r^e{}^2)}}\,.
\eea
Energy conservation then gives
\bea
p_r^{(0)} &=& p_r^{\rm N}\left[1+\eta^2 \left(\frac1{r}-\frac{r}{2a_r^e} \right) \right]
\eea 
with
\beq
p_r^{\rm N}=\sqrt{\frac{2}{r}-\frac{r^2+a_r^e{}^2(1-e_r^e{}^2)}{r^2 a_r^e}}\,,
\eeq
where $e_r^e$ is related to $e_t^e$ by
\beq
e_r^e = e_t^e +3\frac{\eta^2}{a_r^e} e_t^e\,. 
\eeq

One can now compute the averaged value of the first-order Hamiltonian \eqref{H1eob} over a period of the radial motion, Eq. \eqref{Heobaver}, replacing then the EOB orbital parameters $(a_r^e,e_t^e)$ in terms their harmonic coordinate counterparts $(a_r^h,e_t^h)\equiv (a_r,e_t)$ (denoted without the extra label $h$ to ease notation) through the relations
\bea
a_r^e = a_r+\eta^2\,,\nonumber\\
e_t^e = e_t-e_t\nu\frac{\eta^2}{a_r}\,. 
\eea
The averaged Hamiltonian turns out to be
\begin{widetext}
\bea
\langle H^{{\rm tail-of-tail}}_{\rm EOB}\rangle|_{a_r,e_t} &=&\frac{\nu}{a_r^{13/2}}
\left[\frac12 a_{6.5}+\left(\frac{143}{32} a_{6.5}+\frac14 {\bar d}_{5.5}\right) e_t^2\right]\nonumber\\
&+& \frac{\nu\eta^2}{a_r^{15/2}}\left[-\frac{5}{2}a_{6.5}+\frac14 \nu a_{6.5}+\frac12 a_{7.5}\right.\nonumber\\
&+&\left.
\left(-\frac34 {\bar d}_{5.5} -\frac38 \nu {\bar d}_{5.5} -\frac{429}{64}\nu a_{6.5} +\frac{195}{32} a_{7.5}+\frac14 {\bar d}_{6.5}+\frac{411}{32}a_{6.5} \right) e_t^2\right]\,, \nonumber\\
\eea
\end{widetext}
which can then be compared with Eq. \eqref{ris_fin_H}, leading to the final result \eqref{ris_fins}.

\section*{Acknowledgements}

The present research was partially supported by the
2021 Balzan Prize for Gravitation: Physical and Astrophysical
Aspects, awarded to T. Damour.  
D.B. acknowledges 
sponsorship of the Italian Gruppo Nazionale per
la Fisica Matematica (GNFM) of the Istituto Nazionale
di Alta Matematica (INDAM), as well as the hospitality
and the highly stimulating environment of the Institut
des Hautes Etudes Scientifiques.

\end{document}